\documentclass[times,twocolumn]{aastex631}

\usepackage{natbib}
\usepackage{amsmath}
\usepackage{amssymb}
\usepackage{subfigure}
\usepackage{url}
\usepackage{CJK}
\usepackage{longtable}
\usepackage{graphicx}   % import figures
\usepackage{epsfig}
\usepackage{bm}
\usepackage{multirow}

\renewcommand{\vec}[1]{ {\mathbf #1} }

\newcommand{\vol}{ {\mathcal V} }

\newcommand{\RNum}[1]{\uppercase\expandafter{\romannumeral #1\relax}}

\shorttitle{Torus-Unstable Flux Rope Eruption}
\shortauthors{Bian et al.}
\graphicspath{{figures/}}

\begin{document}

\title{Magnetic Reconnection Governs the Entire Lifecycle of Solar Eruptive Magnetic Flux Rope}
	
\correspondingauthor{Chaowei Jiang} \email{chaowei@hit.edu.cn}

\author[0000-0001-9189-1846]{Xinkai Bian}
\affiliation{State Key Laboratory of Solar Activity and Space Weather, School of Aerospace, Harbin Institute of Technology, Shenzhen 518055, China}

\author[0000-0002-7018-6862]{Chaowei Jiang}
\affiliation{State Key Laboratory of Solar Activity and Space Weather, School of Aerospace, Harbin Institute of Technology, Shenzhen 518055, China}
	
\author{Qingjun Liu}
\affiliation{School of Opto-electronic Engineering, Zaozhuang University, Zaozhuang 277160, China}

\author[0000-0001-8605-2159]{Xueshang Feng}
\affiliation{State Key Laboratory of Solar Activity and Space Weather, School of Aerospace, Harbin Institute of Technology, Shenzhen 518055, China}

\author[0000-0003-4711-0306]{Pingbing Zuo}
\affiliation{State Key Laboratory of Solar Activity and Space Weather, School of Aerospace, Harbin Institute of Technology, Shenzhen 518055, China}

\author[0000-0002-7094-9794]{Yi Wang}
\affiliation{State Key Laboratory of Solar Activity and Space Weather, School of Aerospace, Harbin Institute of Technology, Shenzhen 518055, China}

\begin{abstract}

While magnetic flux ropes (MFRs) are widely accepted to exist prior to many solar eruptions, the physical mechanism responsible for triggering and subsequently driving their eruptions remains debated. Classical models commonly attribute eruption onset to ideal magnetohydrodynamic instabilities, particularly the torus instability, while magnetic reconnection is often regarded as a secondary process. Using high-resolution three-dimensional MHD simulations, we investigate the eruption of an MFR that forms self-consistently through flux cancellation driven by converging motions and photospheric diffusion following an initial shearing phase. The pre-eruption MFR rises quasi-statically and reaches a height where the decay index exceeds the classical torus instability threshold, yet the system continues to evolve quasi-statically without eruption.  Instead, eruption begins only when the current sheet beneath the MFR becomes sufficiently thin for magnetic reconnection to set in. Increasing numerical resolution systematically delays the eruption onset, whereas the decay index at eruption onset remains nearly unchanged, demonstrating that the onset is governed by reconnection rather than torus instability. During the eruption, magnetic reconnection continuously converts surrounding sheared arcades into newly reconnected twisted flux that merges with the pre-eruption MFR and continuously to build the erupting MFR. A component-resolved analysis of the Lorentz force work further shows that the newly reconnection flux contributes predominantly to the main acceleration phase. These results demonstrate that magnetic reconnection plays a central role throughout the entire lifecycle of the eruptive MFR: it builds the MFR, directly triggers its eruption, and provides the dominant driving of the eruption. 

\end{abstract}

\section{Introduction} \label{sec:intro}

Coronal mass ejections (CMEs) are the most violent eruptions in the solar system, capable of ejecting magnetized plasma with speeds of hundreds to thousands of kilometers per second and with masses reaching billions of tons. 
These eruptions are powered by free magnetic energy, which is gradually accumulated through photospheric flux emergence and surface motions. 
Prior to eruption, the coronal magnetic field remains in an approximately force-free equilibrium and evolves quasi-statically. 
At a critical point, this force balance is disrupted, leading to rapid magnetic energy release and violent plasma acceleration. 
Two broad classes of mechanisms have been proposed to explain this transition, ideal magnetohydrodynamic (MHD) processes (i.e., ideal instabilities) and non-ideal MHD processes (i.e., magnetic reconnection). 
These two classes of eruption mechanisms have been extensively reviewed~\citep{Forbes2006, Chen2011, Shibata2011, Aulanier2013a, Janvier2015, Patsourakos2020, Jiang2024a}. However, their respective roles in initiating and subsequently driving eruptions remain unresolved.

Magnetic flux ropes (MFRs) are widely regarded as the core magnetic structures of CMEs. An MFR is typically defined as a bundle of magnetic field lines that wind coherently around a common axis. In interplanetary space, magnetic clouds, a subset of interplanetary CMEs, are characterized by enhanced magnetic field strength, a smooth and large-angle rotation of the magnetic field direction, and low plasma $\beta$~\citep{Klein1982}. Magnetic clouds are generally interpreted as the interplanetary manifestations of MFRs erupting from the Sun~\citep{Kunow2006}. In the eruptive source region, an MFR is confined by the overlying strapping field and erupts when the magnetic force balance can no longer be maintained~\citep[i.e., loss of equilibrium,][]{Forbes2000, Lin2000}. Such a loss of equilibrium is equivalent to the well-known torus instability~\citep{Kliem2014}, which occurs when the strapping field decreases sufficiently rapidly with height~\citep{Kliem2006, Demoulin2010}. This interpretation has been supported by many three-dimensional (3D) MHD simulations in which an initially potential bipolar field is energized by photospheric shearing, converging motions, and magnetic flux diffusion, leading to the self-consistent formation and subsequent eruption of a coronal MFR~\citep{Amari2000, Amari2003a, Amari2003, Amari2011a, Aulanier2010, Zuccarello2015}. It was commonly found that in these simulations the decay index of the external field at apex of the MFR axis is very close to the canonical threshold of $n=1.5$ of torus instability, and therefore eruption onset is attributed to a loss of equilibrium or torus instability.

A rising MFR can stretch the magnetic field beneath it and promote the formation and thinning of a current sheet, making magnetic reconnection appear to be a consequence rather than a cause of eruption. In this interpretation, reconnection is regarded primarily as a subsequent flare process that further facilitates the eruption. This causal ordering has encouraged a common practice that uses the decay index of the external strapping field as a primary diagnostic of eruption onset~\citep[e.g.,][]{Cheng2011, Jiang2013a, Amari2018, Inoue2018, Jing2018, Duan2019}, while magnetic reconnection is often regarded as a subsequent process that facilitates the eruption. However, distinguishing the respective roles of torus instability and magnetic reconnection is intrinsically difficult, because the rise of the MFR and magnetic reconnection onset are strongly coupled with each other. As the MFR rises, it may simultaneously compress the overlying field and thereby enhance the downward confinement exerted by the strapping field~\citep{Chen2019}. Conversely, reconnection beneath the MFR can reconfigure the surrounding magnetic field, convert sheared arcades into the twisted MFR, and more importantly, generate a strong upward magnetic tension force at the reconnection outflow region that directly accelerates the erupting structure~\citep{Jiang2021b}. Consequently, finding an MFR in the torus-unstable regime (as commonly determined by the decay index profile) does not by itself establish that the torus instability directly triggers the eruption.

An particularly relevant example was recently provided by \citet{Liu2024d}, who constructed an initially stable pre-eruption MFR using the regularized Biot-Savart law method~\citep{Titov2018, Guo2019a} and drove the system to evolve into an eruption by applying converging motions to the footpoints of the background field. The imposed driving caused the MFR to rise gradually and stretch the overlying field, leading to the formation and progressive thinning of a current sheet beneath it. At the eruption onset, the decay index at the MFR axis was indeed close to the canonical torus instability threshold of $n\approx1.5$, while reconnection also commenced in the current sheet. If only the diagnostic of decay index had been considered, the eruption initiation could therefore have been attributed to the torus instability. However, their simulations performed at different numerical resolutions to resolve the current sheet showed that eruption onset was systematically delayed with increasing resolution, since the onset of reconnection depends on the resolution. This clearly demonstrated that magnetic reconnection in the current sheet beneath the MFR, rather than the torus instability, directly triggered the eruption. Further quantitative analysis showed that the upward magnetic tension force in the newly reconnected flux dominated both the instantaneous upward power and the accumulated work that drove the acceleration of eruption, whereas the pre-eruption MFR contributed only a minor fraction. These results indicate that the decay index alone is insufficient to identify the triggering and driving mechanisms of MFR eruptions and that more comprehensive diagnostics are required. Nevertheless, because the MFR in this model was prescribed in the initial condition, it remains unclear whether the same conclusion applies to a substantial MFR that forms self-consistently before eruption and subsequently enters the torus-unstable regime.

In this work, we extend the study of \citet{Liu2024d} by simulating the full life cycle of an MFR that forms self-consistently from an initially potential bipolar field under photospheric driving until its eruption. An initial shearing phase builds a strongly sheared arcade, after which converging motions and photospheric diffusion transport opposite polarity flux toward the PIL, leading to flux cancellation and the self-consistent formation of a weakly twisted MFR. After the imposed photospheric driving is removed, the MFR subsequently rises quasi-statically and stretches the overlying field, producing a vertical current sheet beneath it, and then an eruption is initiated. Although our simulation is similar in setup to previous simulations such as~\citet{Aulanier2010, Zuccarello2015}, the in-depth analysis following~\citet{Liu2024d} provides a revised interpretation of MFR eruptions, in which magnetic reconnection plays a central role throughout the entire lifecycle of the eruptive MFR: it builds the pre-eruption MFR, triggers its eruption, and provides the dominant driving during its subsequent acceleration.

This paper is organized as follows. Section~\ref{sec:model} describes the model setup. Section~\ref{sec:res} presents the formation, quasi-static rise, and eruption of the MFR, and quantifies the contributions of the original and newly reconnected components of the erupting MFR to the eruption dynamics. Section~\ref{sec:con} presents the discussion and conclusions.

\section{Numerical Model and Setup}\label{sec:model}
We solve the full set of 3D compressible magnetohydrodynamic equations with both coronal plasma pressure and solar gravity included, in a Cartesian coordinate system using the conservation element and solution element (CESE) scheme~\citep{Feng2010, Jiang2010, Jiang2016a, Jiang2021b}. 
The simulations are performed on an adaptive mesh refinement (AMR) grid using the FORTRAN-based \texttt{Paramesh} package~\citep{MacNeice2000}. The overall numerical setup follows our previous work~\citep{Jiang2021b}.
To minimize artificial effects, no explicit resistivity is included in the magnetic induction equation throughout the computational domain, except for the photospheric diffusion term applied at the bottom boundary during the converging phase described below. 
Consequently, in the corona, magnetic reconnection occurs only through numerical diffusion when the thickness of the current layer approaches the local grid scale. This setting ensures that reconnection is not artificially imposed but instead emerges self-consistently from the evolution of the system once a sufficiently thin current sheet has formed.

The MHD equations are solved in normalized form using $B_n=1.86$~G , $L_n=11.52$~Mm, $t_n=105$~s, $v_{n}=110$~km~s$^{-1}$, $\rho_{n}=2.29\times10^{-15}$~g~cm$^{-3}$ and $T_{n}=1\times10^{6}$~K as the normalization constants for magnetic field, length, time, velocity, plasma density and temperature, respectively. Unless otherwise stated, all quantities presented in this paper are converted to physical units, except for time, which is retained in normalized units to facilitate direct comparison among different simulations.
The simulations are carried out in a cubic domain of $[-276,276]\times[-276,276]\times[0,552]$ Mm. The computational grid is block-structured and employs adaptive mesh refinement. The base resolution is $\Delta x=\Delta y=\Delta z=2.88$ Mm. The reference run R0 reaches a finest grid spacing of $180$~km, while the resolution-test runs R1 and R2 reach finest grid spacings of $90$ and $45$~km, respectively. The high resolution is essential for accurately resolving thin current layers and magnetic cancellation near the polarity inversion line.

\begin{figure}[!htb]
	\centering
	\includegraphics[width=1.0\textwidth]{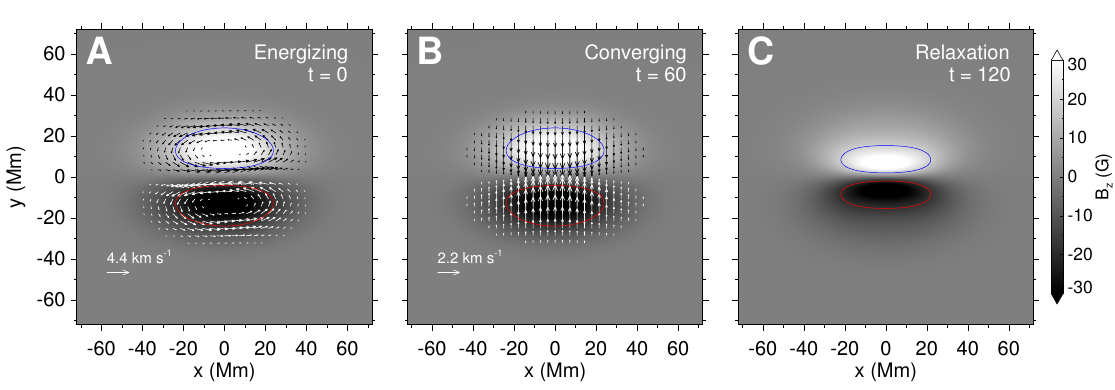}
	\caption{Magnetic flux distribution and surface driving flow at the bottom. (A) Initial magnetic flux distribution and the rotational flow. The background is color-coded by the magnetic normal component $B_z$, and the vectors indicate the imposed rotational flow. 
	The blue and red lines denote one-half of the maximum positive and negative values of the normal magnetic component, respectively. (B) Magnetic flux distribution and the converging flow at $t=60$. (C) The magnetic flux distribution at $t=120$.}\label{fig:map}
\end{figure}

The initial background atmosphere is stratified by solar gravity, with a bottom density of $\rho=2.29\times10^{-15}$~g~cm$^{-3}$ and an uniform temperature of $T=10^6$~K.
The initial magnetic field is a potential field computed from a prescribed bipolar photospheric magnetogram using the Green's function method. The vertical magnetic field component is represented by two Gaussian functions~\citep{Amari2003a}:
\begin{equation}\label{eq:magnetogram}
	\begin{split}
		B_{z}(x,y,0) = B_0
		e^{-x^2/\sigma_x^2}(e^{-(y-y_c)^2/\sigma_y^2}-e^{-(y+y_c)^2/\sigma_y^2}).
	\end{split}
\end{equation}
Here, $B_0=37.2$~G. The parameters $\sigma_x=28.8$~Mm and $\sigma_y=14.4$~Mm define the spatial extent of the magnetic polarities in the $x-$ and $y-$ directions, and $y_c=11.52$~Mm specifies the separation distance between the centers of the opposite magnetic polarities. The magnetogram is shown in Fig.~\ref{fig:map}A.

To mimic the formation of an MFR in solar active regions, we impose rotational and converging flows together with photospheric flux diffusion at the bottom boundary. 
In the first stage, a rotational flow is applied at the bottom boundary to inject magnetic free energy and build up magnetic shear along the PIL. The velocity field is defined as,
\begin{equation}\label{eq:dirven_speed}
	\begin{split}
		v_{x}=\dfrac{\partial \psi(B_{z})}{\partial y};
		v_{y}=-\dfrac{\partial \psi(B_{z})}{\partial x},
	\end{split}
\end{equation}
with 
\begin{equation}\label{eq:dirven_speed_psi}
	\begin{split}
		\psi = v_{0}B_{z}^{2}e^{-(B_{z}^{2}-B_{z, {\rm max}}^{2})/B_{z, {\rm max}}^{2}}.
	\end{split}
\end{equation}
This driving follows the contour of $B_z$, thereby preserving the magnetic flux distribution while injecting magnetic shear. Here, $v_0$ is chosen such that the maximum driving speed is $4.4$~km~s$^{-1}$ (i.e., $0.04 v_n$), much smaller than the typical Alfv$\acute{\text{e}}$nic velocity ($\sim1600$~km~s$^{-1}$) in the bottom magnetic field region, thereby ensuring quasi-static evolution.

%After a significant delay, long before the eruption occurs, the rotational driving is halted, and a converging driving along with flux diffusion at the photosphere are introduced to facilitate the formation and elevation of the MFR. Once the MFR reaches a certain height, all driving are halted, allowing the system to evolve spontaneously to explore the role of the MFR in the eruption initiation process.
%
%
%In our simulation, the first rotational driving motion mimics the rotational motion of sunspots and is implemented through a velocity field derived from the magnetic flux distribution at the bottom boundary. This velocity effectively injects free magnetic energy into the system while without modifying the magnetic flux distribution. The velocity are defined as follows,

At $t=60$, well before the onset of eruption, the rotational driving is switched off and replaced by a converging flow combined with surface flux diffusion. 
This transition marks the beginning of the converging phase, during which opposite polarity flux is transported toward the PIL, where flux cancellation is facilitated by the combined effects of the converging motion and photospheric diffusion. The converging motion is given by,
\begin{equation}\label{eq:converg}
	\begin{split}
		v_x = 0; v_y = -v_1 B_z(t=0),
	\end{split}
\end{equation}
where $v_1$ is also a constant for scaling such that the maximum
velocity is $2.2$~km~s$^{-1}$ (i.e., $0.02 v_n$). The maximum velocities adopted for the shearing and converging phases, $4.4$ and $2.2$~km~s$^{-1}$, respectively, were chosen as a compromise between	observational realism and computational efficiency. While these values are larger than typical photospheric horizontal velocities of order $1$~km~s$^{-1}$ or less, using substantially smaller speeds would	increase the computational cost, whereas much faster driving could	compromise the quasi-static evolution of the coronal magnetic field. The lower speed adopted during the converging phase also helps avoid unrealistically rapid flux convergence and cancellation.
During the converging process, we follow previous studies \citep{Aulanier2010, Rice2023} by introducing a photospheric diffusion coefficient into the magnetic induction equation to model photospheric flux cancellation,
\begin{equation}\label{eq:mag_induc}
	\begin{split}
	\dfrac{\partial \vec{B}}{\partial t}= \nabla\times(\vec{v} \times \vec{B}) + \eta_{\text{pho}} \nabla_{\perp}^2 \vec{B},
    \end{split}
\end{equation}
where $\eta_{\text{pho}}=2\times 10^{-3}$. This corresponds to a characteristic diffusion timescale of $500 t_n$ ($5.25\times10^4$~s), which is much longer than the typical Alfv$\acute{\text{e}}$nic timescale of $\sim6.3$~s in the bottom magnetic field region, corresponding to local Alfv$\acute{\text{e}}$nic speeds exceeding $1600~{\rm km~s^{-1}}$. 
The relatively small driving speed and diffusion coefficient ensure that the system evolves quasi-statically throughout this phase. The imposed converging flow is shown in Fig.~\ref{fig:map}B. After a driving duration of $\Delta t=60$, the magnetic flux becomes significantly more compact near the PIL, as shown in Fig.~\ref{fig:map}C. During this phase, as facilitated by the photospheric diffusion, slow reconnection between sheared arcades occurs at the PIL, which gradually builds up an MFR.

After the MFR is formed, all imposed photospheric driving is switched off. The bottom boundary remains line-tied ($\mathbf{v}=0$), so that the subsequent evolution is governed by the coronal magnetic field itself, without further photospheric driving.
This setup enables us to isolate the role of the pre-eruption MFR and determine whether it can trigger eruption in the absence of further photospheric driving.

\section{Results}\label{sec:res}

\subsection{Formation and Quasi-static Rise of the Magnetic Flux Rope}\label{sec:form}

\begin{figure}[!htb]
	\centering
	\includegraphics[width=1.0\textwidth]{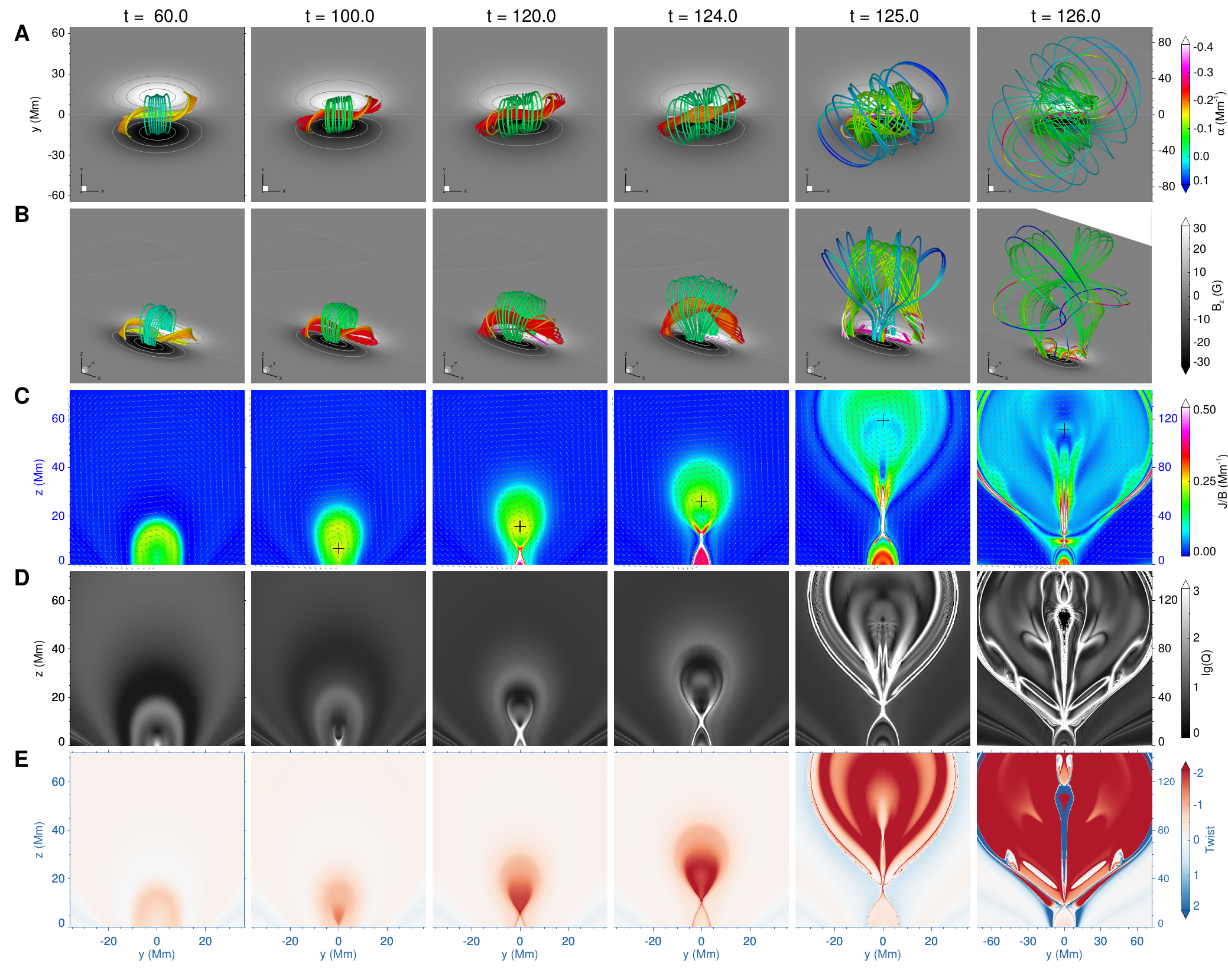}
	\caption{Evolution of magnetic field, normalized current, plasma velocity, magnetic squashing factor Q and twist in the simulation. (A) Two-dimensional view of selected magnetic field lines. The selected magnetic field lines are colored according to the force-free factor $\alpha=\vec{J}\cdot\vec{B}/B^2$. 
	The background shows the magnetic flux distribution on the bottom boundary. (B) 3D perspective view of the same field lines shown in panel (A). (C)–(F) Distributions of the normalized current density $J/B$, plasma velocity magnitude, magnetic squashing factor $Q$, and magnetic twist number $T_w$, respectively, on the central vertical slice. The arrows in panel (C) represent the normalized transverse magnetic field vectors on the slice, revealing a spiral pattern from which the MFR axis is identified and marked by a plus sign. The panels at $t=126$ are shown with a larger field of view.
	An animation of this figure is available. In the animation, panels (A) and (B) are displayed on the left, panel (C) is shown in the upper middle without the normalized transverse magnetic field component, and the plasma velocity distribution on the central vertical slice is shown in the lower middle. Panels (D) and (E) are displayed on the right.} \label{fig:mag}
\end{figure}

\begin{figure}[!htb]
	\centering
	\includegraphics[width=0.7\textwidth]{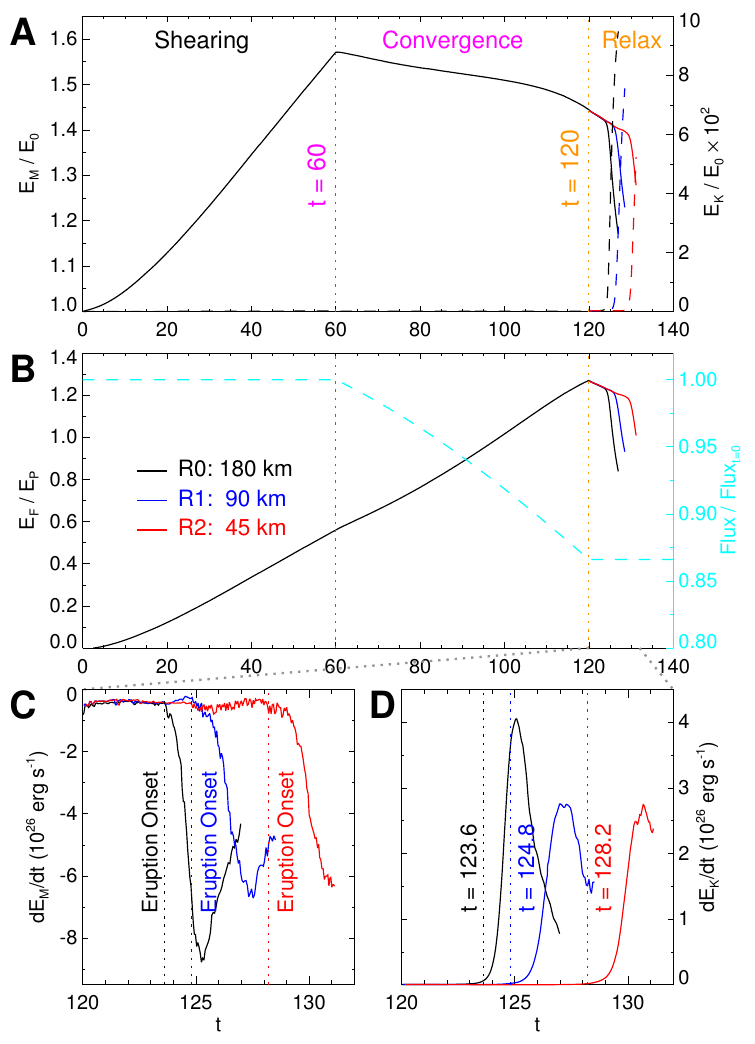}
	\caption{Temporal evolution of magnetic energy, kinetic energy, and magnetic flux. (A) Temporal evolution of the magnetic energy during different driving stages. The black, magenta, and orange shaded regions denote the shearing, converging (flux diffusion), and relaxation stages, respectively. The black, blue, and red solid lines correspond to simulations with the finest grid spacings of $180$, $90$, and $45$ km, respectively, while the dashed lines show the corresponding kinetic energies. The vertical dotted lines mark the transitions between different driving stages.
	(B) Evolution of the non-potentiality ($E_F/E_P$) and the total unsigned magnetic flux. (C) and (D) Time derivatives of the magnetic and kinetic energies. 
    The $E_M$, $E_P$, $E_F$, $E_0$ and $E_K$ refer to the magnetic energy within the system, the potential field energy (varying with the magnetic flux distribution), the free magnetic energy (magnetic energy minus potential field energy),  the potential field energy at $t=0$ and the kinetic energy, respectively. The initial potential-field energy is $E_0 = 7.32\times10^{29}$ erg. The vertical dotted lines indicate the eruption onset.}\label{fig:energy}
\end{figure}

The simulation begins with a potential magnetic field subjected to photospheric shearing motion (Fig.~\ref{fig:map}A). The imposed rotational flow injects magnetic free energy and progressively builds up strong shear along the PIL. 
By $t=60$, the low-lying field lines in the core region are strongly sheared and become nearly aligned with the PIL (Fig.~\ref{fig:mag}A), while the central region of each polarity remains close to a potential configuration. 
To quantify the accumulated twist of the flux around the center of the polarities, we estimate the averaged displacement of the field-line footpoints along the $|B_z|=0.5B_{z,\max}$ contour. Over the shearing phase ($t=0$--$60$), the accumulated displacement is approximately $0.13$ times the contour length, indicating that the flux connecting the two polarities is twisted by around $0.26$ turns. 
Throughout this phase, the magnetic energy increases monotonically without any significant rise in kinetic energy (Fig.~\ref{fig:energy}A), indicating that the system evolves quasi-statically.

At $t=60$, the shearing motion is switched off and replaced by converging motion combined with photospheric flux diffusion. 
This driving transports opposite magnetic polarities toward the PIL, leading to sustained magnetic flux cancellation. 
The distribution of the magnetic squashing factor $Q$ reveals the formation of quasi-separatrix layers (QSLs) above the PIL, which outline the evolving magnetic topology (Fig.~\ref{fig:mag}E).
As flux cancellation proceeds, reconnection between sheared arcade field lines across the PIL gradually builds up a weakly twisted MFR above the PIL (Fig.~\ref{fig:mag}B). 
This demonstrates that the MFR forms self-consistently through reconnection driven by flux cancellation, rather than being imposed as an initial condition. 
During this stage, the total magnetic energy decreases gradually, whereas the degree of non-potentiality, defined as the ratio of free magnetic energy to potential field energy, continues to increase (Fig.~\ref{fig:energy}B).

As the MFR accumulates magnetic flux, it gradually separates from the photosphere and rises slowly. 
During this process, the magnetic topology evolves from a bald-patch separatrix surface~\citep[BPSS,][]{Gibson2004, Green2009, Jiang2013}, in which the lower boundary of the MFR remains attached to the photosphere, into a hyperbolic flux tube~\citep[HFT,][]{Titov2002, Aulanier2005}, characterized by an X-type topology beneath the MFR in the central vertical slice (Fig.~\ref{fig:mag}D). 
This BPSS-to-HFT transition indicates that the body of the MFR has become fully separated from the photosphere and that a vertical current sheet has formed beneath it.
By $t=120$, the HFT is fully established, clearly separating the MFR from the surrounding sheared arcades. During the converging phase, the unsigned magnetic flux decreases by approximately 13$\%$ as a result of the sustained flux cancellation. 
At $t=120$, all imposed photospheric driving is switched off instantly without producing any noticeable disturbance in the subsequent evolution of the MHD system, while the bottom boundary remains line-tied with zero velocity. 
The system is thereafter allowed to evolve solely under the action of the coronal magnetic field, enabling us to isolate the role of the pre-eruption MFR without further photospheric driving.

Following the definition of an MFR as a bundle of magnetic field lines winding around a common axis, we identify the MFR axis on the central vertical slice as the center of spiral pattern formed by the normalized transverse magnetic field, as shown in Fig.~\ref{fig:mag}C. 
The apex height of the MFR axis is tracked throughout the simulation (Fig.~\ref{fig:nindex}A). 
During the relaxation stage, the MFR continues to rise quasi-statically, with an apex speed of only $\sim5$~km~s$^{-1}$, increasing modestly to $\sim10$~km~s$^{-1}$ immediately before eruption. Meanwhile, the magnetic energy decreases gradually without any impulsive release (Fig.~\ref{fig:energy}), confirming that the system remains in a quasi-static state despite the continued slow expansion of the MFR. 
This quasi-static evolution persists until the abrupt transition to eruption at $t=123.6$, which is analyzed in the following subsection.

\subsection{Eruption Onset Triggered by Magnetic Reconnection}\label{sec:reconnect}
\begin{figure}[!htb]
	\centering
	\includegraphics[width=0.9\textwidth]{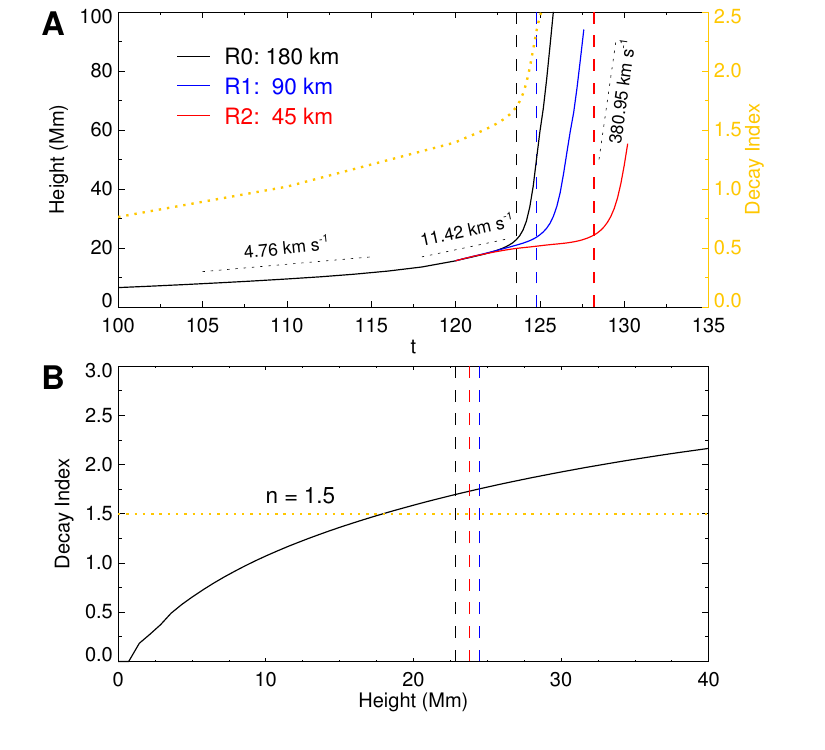}
	\caption{Evolution of the MFR axis and the decay index. (A) Apex height (solid lines) of the MFR axis and the decay index (dashed lines) of the strapping field at the apex of the MFR axis. Black, blue, and red correspond to runs R0, R1, and R2, respectively. 
	The vertical dash–dotted lines denote the onset times of the eruptions. (B) Decay index as a function of height. The horizontal dashed line marks the critical decay index value of $n=1.5$.}\label{fig:nindex}
\end{figure}

The system evolves quasi-statically until $t=123.6$, when it undergoes an abrupt transition into the eruptive phase. 
As shown in Fig.~\ref{fig:energy}, this transition is characterized by a rapid decrease in magnetic energy accompanied by a sharp increase in kinetic energy, indicating an efficient conversion of magnetic energy into plasma kinetic energy. At the eruption onset, the non-potentiality reaches a degree of approximately $1.2$ (Fig.~\ref{fig:energy}B), indicating that the magnetic field has accumulated a substantial amount of free energy prior to eruption.
Subsequently, the magnetic energy decreases by approximately $0.25E_0$ (where $E_0 = 7.32\times10^{29}$ erg is the potential field energy at $t=0$), while the kinetic energy increases by about $0.10E_0$. At the same time, the MFR axis departs from its previous quasi-static evolution and begins an exponential rise (Fig.~\ref{fig:nindex}A), with its rising speed increasing rapidly to approximately $380$~km~s$^{-1}$. Meanwhile, the MFR expands rapidly as it erupts outward (as can be seen in the last column of Fig.~\ref{fig:mag}). We therefore define $t=123.6$ as the eruption onset.

To identify the physical mechanism responsible for this transition, we first examine the state of the system immediately before eruption onset. In the present bipolar configuration, the overlying magnetic field, represented by the $y$-component ($B_y$), provides the dominant downward confinement of the MFR (Fig.~\ref{fig:mag}A). Accordingly, we evaluate the decay index of the overlying field, $n=-{\rm d~}{\rm ln}(-B_{y})/{\rm d~}{\rm ln}(z)$, at the apex of the MFR axis (Fig.~\ref{fig:nindex}). 
The MFR axis reaches a height corresponding to a decay index of $n\approx1.7$, exceeding the canonical threshold for torus instability. However, after entering the torus-unstable regime, the system continues to evolve quasi-statically until $t=123.6$, without immediately erupting. 
This result demonstrates that satisfying the torus instability criterion alone is insufficient to initiate eruption.

\begin{figure}[!htb]
	\centering
	\includegraphics[width=0.9\textwidth]{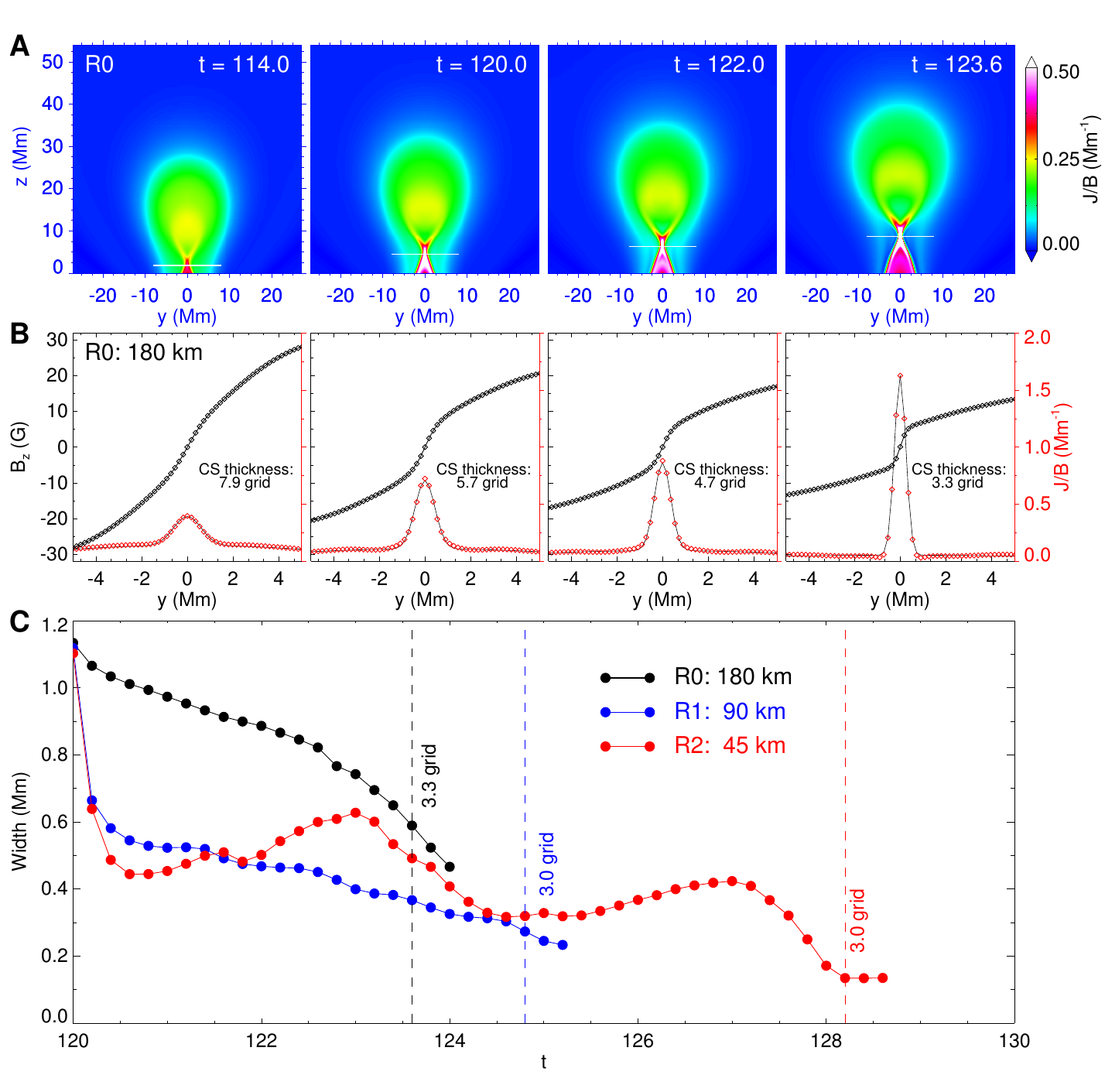}
	\caption{Formation and evolution of the core current sheet. (A) Distribution of normalized current $J/B$ on the $x=0$ slice in simulation run R0. (B) One-dimensional profile of the magnetic component $B_z$ and normalized current density along the horizontal line perpendicular to the center of the core CS (location with the maximum $J/B$). The diamonds represent the value on the grid nodes. The thin black line is a Gaussian fit of the normalized current density $J/B$ profile, and its FWHM is used to measure the thickness
	of the CS. (C) Temporal evolution of the thickness of the core CS in three simulation runs. Circles indicate individual simulation outputs.}\label{fig:jc}
\end{figure}

As the MFR continues to rise slowly, it progressively stretches the overlying magnetic field and continuously thins the current layer beneath it (Fig.~\ref{fig:jc}). 
The thickness of current layer decreases steadily until it approaches the local grid scale ($\sim3$--$4$ grid cells), at which point magnetic reconnection sets in through numerical diffusion. 
Remarkably, the onset of magnetic reconnection coincides precisely with the eruption onset defined above, including the rapid release of magnetic energy, the impulsive increase of kinetic energy, and the onset of the exponential rise of the MFR axis.
The coincidence of these signatures demonstrates that magnetic reconnection marks the transition from the quasi-static evolution to the eruptive phase. 
Therefore, although the MFR has already entered the torus-unstable regime, eruption does not commence until magnetic reconnection is initiated beneath the MFR, indicating that magnetic reconnection, rather than the torus instability, serves as the direct trigger of the eruption.

\subsection{Resolution Dependence of Eruption Onset} \label{sec:resolu}
If the eruption were triggered solely by an ideal MHD instability, such as the torus instability, its onset should be independent of numerical resolution. 
To test this hypothesis, we restart the simulation from the common state at $t=120$, when the pre-eruption MFR has formed and all imposed photospheric driving has been halted. Three simulations are then performed with finest grid spacings of $\Delta=180$~km, $90$~km, $45$~km, hereafter referred to as R0, R1 and R2, respectively, with R0 serving as the reference run. In this way, numerical resolution is the only parameter that differs among the three simulations.

The results show that the eruption onset is systematically delayed with increasing numerical resolution (Fig.~\ref{fig:energy}). In contrast, both the apex height of the MFR axis and the corresponding decay index at eruption onset remain nearly unchanged among the three runs, with $n \approx 1.7$ in all cases (Fig.~\ref{fig:nindex}).
Thus, each simulation reaches essentially the same torus-unstable state before eruption, whereas the eruption onset time depends strongly on numerical resolution.

The evolution of the current sheet thickness explains this behavior (Fig.~\ref{fig:jc}C). At higher resolution, the current sheet continues to thin to smaller spatial scales before numerical diffusion becomes effective and reconnection sets in. Consequently, both the onset of reconnection and the subsequent eruption are systematically delayed.
This resolution dependence demonstrates that the eruption onset is not governed by an ideal MHD instability, but instead by the formation of a sufficiently thin current sheet capable of initiating magnetic reconnection. 
Since the onset of an ideal MHD instability should be independent of numerical resolution, these results provide strong evidence that magnetic reconnection, rather than the torus instability alone, directly triggers the eruption.

\subsection{Magnetic Reconnection Dominates the Eruption Dynamics}\label{sec:work}
Having established that magnetic reconnection triggers the eruption, we next investigate whether it also provides the dominant driving force during the eruption. 
To this end, we separate the entire erupting MFR into its pre-eruption and newly reconnected components using a field-line tracing method similar to that of \citet{Liu2024d}.

\begin{figure}[!htb]
	\centering
	\includegraphics[width=0.9\textwidth]{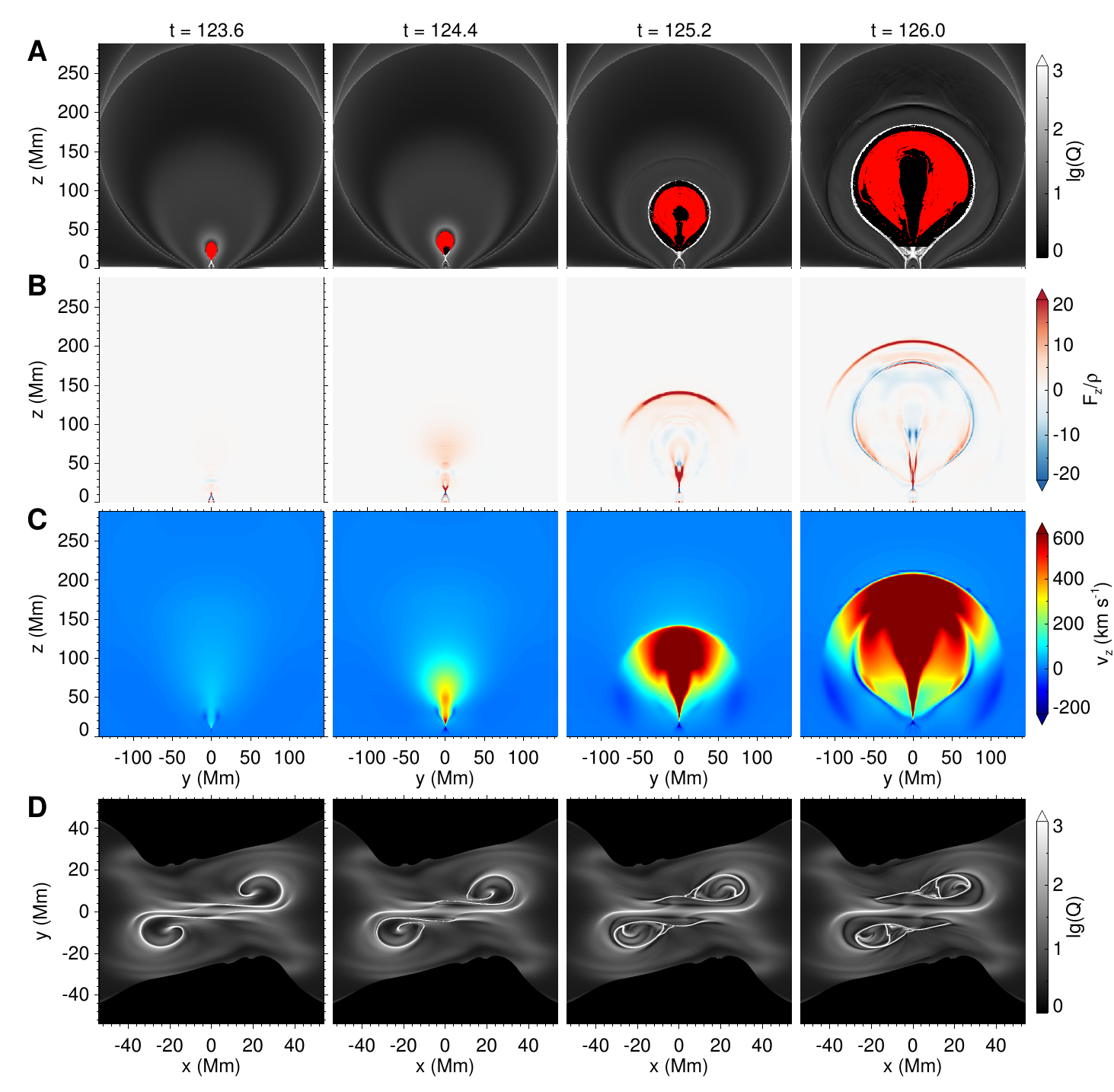}
	\caption{Evolution of the erupting MFR on the central vertical slice. (A) Distribution of the magnetic squashing factor $Q$ on the central vertical slice. The red region outlines the cross section of the pre-eruption MFR identified by the high-$Q$ QSLs. 
	(B) Distribution of the vertical Lorentz-force acceleration, $F_z/\rho$. (C) Distribution of the vertical velocity $v_z$. Snapshots are taken from eruption onset to the end of the main acceleration phase. (D) Distribution of the magnetic squashing factor $Q$ on the bottom boundary, i.e, $z=0$. 
	}\label{fig:mfr}
\end{figure}

First, at the eruption onset ($t=123.6$), we identify the pre-eruption MFR using the magnetic squashing factor $Q$. Specifically, the closed region enclosed by high-$Q$ QSLs in the central vertical slice is taken as the boundary of the pre-eruption MFR (Fig.~\ref{fig:mfr}A). Magnetic field lines rooted within this region are traced, and their footpoints on the bottom boundary are recorded. These footpoints define the footprint of the pre-eruption MFR on the bottom boundary.
At subsequent times during the eruption, we identify the entire erupting MFR by selecting field-line seed points inside the QSL-enclosed erupting region which expands with time. 
For each traced field line, both footpoints are mapped to the bottom boundary and compared with the pre-eruption MFR footprint defined at $t_{\rm onset}$. Only field lines whose two footpoints both belong to the pre-eruption MFR footprint are classified as the original component that is formed before the eruption onset. All remaining field lines are classified as the newly reconnected component that is created during the eruption. This procedure allows us to reconstruct the full 3D structure of the erupting MFR and separate it into original and newly reconnected components.

The photospheric distributions of $Q$ further show the evolution of the footpoint regions of the magnetic flux involved in reconnection during the eruption (Fig.~\ref{fig:mfr}D). At eruption onset, the high-$Q$ footprints exhibit a characteristic double-J shaped pattern on the two sides of the PIL. As the eruption proceeds, the hooked parts of the J-shaped structures progressively become closed, while the high-$Q$ footprints gradually move away from the PIL. These changes illustrate the evolution of the photospheric footpoint regions associated with the magnetic flux participating in the reconnection process.

Using this classification, we further analyze the eruption dynamics. Fig.~\ref{fig:mfr}B-C shows the distributions of the vertical acceleration $F_z/\rho$ by Lorentz force and vertical velocity $v_z$ on the central cross-section, where $F_z = (\vec{J} \times \vec{B})_z$. 
Strong upward acceleration is concentrated immediately above the current sheet, where the reconnection outflow is formed, and the upward velocity exceeds $600$~km~s$^{-1}$ in this region.

To quantify which component dominates the upward driving of the eruption, we calculate the upward power and accumulated work done by the Lorentz force for each component. The upward power is defined as
\begin{equation}\label{eq:p}
	\begin{split}
		P_s(t)&=\int_{\mathcal{V}_s ,F_z>0,\,v_z>0} F_z v_z\,{\rm d}\vol_s,
	\end{split}
\end{equation}
where the subscript $s$ denotes the entire erupting MFR ($e$), the pre-eruption (original) component ($o$), and the newly reconnected component ($r$). The integral is evaluated over the 3D volume $\mathcal{V}_s$ occupied by component $s$. The quantity therefore measures the instantaneous rate at which the vertical Lorentz force does positive work to accelerate plasma upward. 
We evaluate the contributions from all regions within $\mathcal{V}_s$ that are excluded by Equation~(\ref{eq:p}). Their combined contribution is typically about $5\%$ of the corresponding full volume contribution in $\mathcal{V}_s$ during the eruption. Therefore, the excluded contributions are overall small, and Equation~(\ref{eq:p}) captures the dominant Lorentz force power associated with the eruption.
The accumulated work is given by
\begin{equation}\label{eq:w}
	\begin{split}
		W_s(t)&=\int_{t_{\rm onset}}^{t} P_s (t') \text{d}t',
	\end{split}
\end{equation}
which measures the accumulated contribution of each component to the upward acceleration of the eruption. 
Accordingly, the calculated power and work quantify the upward driving of the erupting MFR.

\begin{figure}[!htb]
	\centering
	\includegraphics[width=0.9\textwidth]{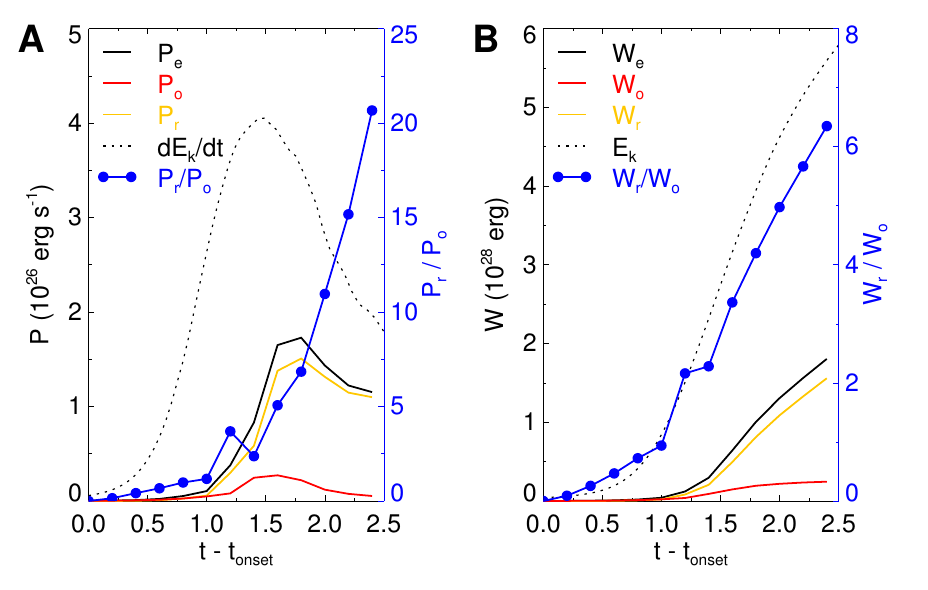}
	\caption{Evolution of the upward power and accumulated work associated with the erupting MFR. The time axis is shifted so that $t=0$ corresponds to the eruption onset at $t=123.6$ in the simulation.	 
	(A) The black, red, and orange solid lines represent the upward power associated with the entire erupting MFR, original component of the erupting MFR, and reconnected component of the MFR, respectively. The black dashed line shows the growth rate of the total kinetic energy. The blue curve shows the ratio of the upward power of the newly reconnected component to that of the original component. 
	Each circle represents a moment of output data. 
	(B) Same as panel (A), except that the curves represent the accumulated work.}\label{fig:pe}
\end{figure}

Figure~\ref{fig:pe}A shows the temporal evolution of the upward power associated with the erupting MFR. The black, red, and orange solid lines represent the entire erupting MFR, the original, and the newly reconnected component, respectively. 
For comparison, the black dashed line shows the growth rate of the total kinetic energy. 
Hereafter, $\delta t=t-t_{\rm onset}$ denotes the elapsed time since eruption onset, with $\delta t=0$ corresponding to $t_{\rm onset}=123.6$ in the simulation.
Following eruption onset, all power curves increase rapidly, while the kinetic energy growth rate reaches a maximum at $\delta t=1.5$ and subsequently decreases to approximately half of its peak value by $\delta t=2.5$. We define this interval as the main acceleration phase of the eruption.

During this phase, the upward power of the original component reaches its maximum earlier, at $\delta t=1.5$, whereas both the entire erupting MFR and the newly reconnected component peak slightly later, at $\delta t=1.8$. The upward power of the entire erupting MFR is approximately one-half of the total kinetic energy growth rate. 
This difference is expected because the kinetic energy growth rate includes plasma motions in all directions throughout the computational domain, whereas the power defined here includes only the positive work done by the vertical Lorentz force within the MFR volume limited by the upward motion. 
Moreover, plasma outside the QSL-enclosed region also contributes to the increase in the total kinetic energy. 
Importantly, the ratio of the upward power contributed by the newly reconnected component to that of the original component increases continuously throughout the main acceleration phase, reaching approximately $21$ at $\delta t=2.5$ (blue curve in Fig.~\ref{fig:pe}A). This demonstrates that the newly reconnected component dominates the instantaneous upward driving of the eruption.

Figure~\ref{fig:pe}B presents the accumulated upward work done by the Lorentz force. Because the upward power remains positive throughout the main acceleration phase, the accumulated work increases monotonically with time. By $\delta t=2.5$, the total upward work associated with the erupting MFR reaches approximately one-third of the total increase in kinetic energy, again reflecting the fact that only the vertical contribution within the MFR volume is included. 
The ratio of the accumulated work contributed by the newly reconnected component to that of the original component reaches approximately $6.4$ at $\delta t\approx2.5$. This corresponds to a contribution of about $6.4/(1+6.4)\approx86.5\%$ of the total upward work. 
Therefore, the newly reconnected component dominates not only the instantaneous upward power but also the accumulated upward work throughout the main acceleration phase. 
%These results provide direct quantitative evidence that magnetic reconnection not only triggers the eruption, but also continuously builds the erupting MFR and provides the dominant driving mechanism throughout its main acceleration phase.

Taken together with the results of Section~\ref{sec:reconnect}, these analyses demonstrate that magnetic reconnection governs the entire eruption process. It not only initiates the eruption, but also continuously builds the erupting MFR and provides the dominant upward driving throughout the main acceleration phase.

\section{Discussion}  \label{sec:con}

In many previous MHD simulations such as \citet{Amari2000, Amari2003a, Amari2003}, \citet{Aulanier2010}, and \citet{Zuccarello2015} that described the full life cycle of coronal eruptive MFRs from birth to eruption, the initiation mechanism is commonly attributed to torus instability~\citep{Kliem2006}, according to the result that the decay index approached the canonical threshold of $n\approx1.5$ at eruption onset. Magnetic reconnection beneath the rising MFR was therefore interpreted mainly as a consequence of eruption or as a secondary process that facilitated its subsequent development. 

In this work, we conducted a similar simulation while offering new insights on the trigger and driver of MFR eruption. First, although the MFR rises quasi-statically into the torus-unstable regime, no eruption occurs at this stage, demonstrating that satisfying the torus-instability criterion alone is insufficient to initiate eruption. Second, eruption begins only when the underlying current sheet becomes sufficiently thin for magnetic reconnection to set in, as strongly supported by the systematic delay of eruption onset with increasing numerical resolution. Third, following eruption onset, magnetic reconnection continuously adding new flux to the erupting MFR and provides the dominant upward driving during the main acceleration phase, with the newly reconnected flux contributing approximately nearly $90\%$ of the total upward work done by the Lorentz force. The resolution dependent eruption onset and component-resolved work analysis therefore identify magnetic reconnection as the direct trigger and dominant driver of the eruption. Magnetic reconnection thus plays a central role throughout the entire lifecycle of the eruptive MFR: it builds the pre-eruption MFR, triggers its eruption, and provides the dominant driving during its acceleration. These results further indicate that decay index diagnostic alone is insufficient to identify the eruption mechanism and should be complemented by analyses of current sheet evolution, numerical resolution dependence, reconnection onset, and force or work decomposition.

The present simulation is performed with a simple bipolar configuration, therefore only a current sheet is formed and reconnection occurs right beneath the MFR. When the magnetic configuration is complex, reconnection may occur at different locations and affect the eruption in different ways. For example, in a quadrupolar configuration, reconnection can occur between the rising MFR and its overlying field, and facilitates eruption by removing the overlying confining flux~\citep{Antiochos1999}. However, when the ambient field is sufficiently strong and its orientation favors sustained reconnection with the MFR, such external reconnection can progressively erode the MFR and ultimately lead to a failed eruption~\citep{Zhou2022, Jiang2023, Chen2024, Guo2024}. Thus, whether magnetic reconnection facilitates or suppresses an eruption depends on its location, the relative orientation of the interacting magnetic fields, and the amount of magnetic flux involved, which requires further studies.

%One limitation of the present quantitative analysis is that the pre-eruption and newly reconnected components are identified solely on the basis of their photospheric footpoint connectivity. 
%In reality, magnetic reconnection within the erupting MFR, including rope--rope reconnection, may occur during the eruption and further modify the internal connectivity of the MFR~\citep{Aulanier2019}. 
%Such reconnection is likely to occur in a erupting system and may increase the twist of the erupting MFR. 
%Nevertheless, this effect is unlikely to alter our main conclusions for two reasons. 
%First, the pre-eruption MFR expands substantially during the eruption, and our field-line tracing method likely provides an upper estimate of the spatial extent of the original component. 
%Second, the strongest upward acceleration occurs in the reconnection outflow region above the current sheet, where sheared arcade flux is converted into newly reconnected flux. 
%Therefore, the dominant contribution to the upward power and accumulated work is robustly associated with the reconnection-formed component. 
%Future work incorporating time-dependent magnetic connectivity tracking will further quantify the role of reconnection within the erupting MFR during the later stages of eruption.

\begin{acknowledgments}
  \modulolinenumbers[10]
  This work is jointly supported by National Natural Science Foundation of China (NSFC 12573058, 42504160), Shenzhen Science and Technology Program (grant no. RCJC20210609104422048), Guangdong Basic and Applied Basic Research Foundation (2023B1515040021) and the Fundamental Research Funds for the Central Universities (HITJCRC2026005). The computational work was carried out on TianHe-1(A), National Supercomputer Center in Tianjin, China.
\end{acknowledgments}

\bibliography{mylib}
\bibliographystyle{aasjournal}

%\begin{figure}[!htb]
%	\centering
%	\includegraphics[width=0.9\textwidth]{para_evol_no_e_0120.pdf}
%	\caption{Temporal evolution of magnetic and kinetic energy in no eruption case. The lines are the same as in Fig.~\ref{fig:energy}A, except that the driving stop time is $t=110$.}\label{fig:energy_no_e}
%\end{figure}

\end{document}